\documentclass[11pt]{article}
\usepackage[margin=1 in,top=1 in,bottom=1 in]{geometry}
\usepackage[utf8]{inputenc}
\usepackage{graphicx}
\usepackage{amssymb}
\usepackage{amsmath}
\usepackage{float}
\usepackage{subfigure}
\usepackage{authblk}
\usepackage{soul}
\usepackage[linktocpage=true,colorlinks=true,citecolor=blue,linkcolor=blue,urlcolor=blue]{hyperref}

\begin{document}
\title{Surface and Bulk Relaxation of Vapour-Deposited Polystyrene Glasses}
\author[1]{Junjie Yin}
\author[2]{Christian Pedersen}
\author[1]{Michael F. Thees}
\author[2]{Andreas Carlson}
\author[3]{Thomas Salez}
\author[1]{James A. Forrest}
\affil[1]{Department of Physics \& Astronomy, University of Waterloo, 200 University Ave. W, Waterloo, ON, N2L 3G1, Canada }
\affil[2]{Mechanics Division, Department of Mathematics, University of Oslo, 0316 Oslo, Norway}
\affil[3]{Univ. Bordeaux, CNRS, LOMA, UMR 5798, F-33400, Talence, France}
\maketitle

\abstract{We have studied the liquid-like response of the surface of vapour-deposited glassy films of polystyrene to the introduction of gold nanoparticles on the surface. The build-up of polymer material was measured as a function of time and temperature for both as-deposited films, as well as films that have been rejuvenated to become normal  glasses cooled from the equilibrium liquid. The temporal evolution of the surface profile is well described by the characteristic power law of capillary-driven surface flows. In all cases, the surface evolution of the as-deposited films and the rejuvenated films are enhanced compared to bulk and are not easily distinguishable from each other. The temperature dependence of the measured relaxation times determined from the  surface  evolution is found to be quantitatively comparable to similar studies for high molecular weight spincast polystyrene. Comparisons to numerical solutions of the glassy thin film equation provide quantitative estimates of the surface mobility. For temperatures sufficiently close to the glass-transition temperature, particle embedding is also measured and used as a probe of bulk dynamics, and in particular bulk viscosity.}

\section{Introduction}
The surface dynamics of glasses is a key area of study in condensed matter and materials physics \cite{tian2022surface}. In particular, it has become increasingly evident that the first few nanometers of a glassy material can exhibit properties vastly different from the bulk of the glass. In addition to its key role in the ability to form what have been called ultrastable glasses \cite{berthier2017origin}, the surface mobility is widely believed to be the key underlying factor in the 25 year long study of reduced glass transition temperatures $T_{\textrm{g}}$ in thin polymer films \cite{sharp2003free,forrest2014does, ediger2014dynamics}.

Since their discovery 15 years ago \cite{swallen2007organic}, kinetically stable glasses produced by vapour deposition have been extensively studied. While most studies have involved molecular glasses, more recent studies have included metallic \cite{luo2018ultrastable}, as well as polymeric glasses \cite{raegen2020ultrastable}. Collectively, these materials have strong similarities to  glasses aged to near equilibrium. Many vapour deposited materials have been shown to have lower entropy and be lower in the energy landscape than glasses formed by cooling from the equilibrium supercooled liquid \cite{ramos2011character, leon2010stability, leon2016correction}. In fact, recent studies on vapour-deposited polymer glasses have suggested a similarity to materials aged for as much as $10^{13}$ s \cite{raegen2020ultrastable}. These unique properties are proposed to result from both the enhanced mobility of glassy surfaces, and the layer-by-layer formation associated with the method of vapour deposition~\cite{berthier2017origin}. While it might be expected that as-deposited stable films and rejuvenated glassy films both have enhanced surface mobilities, it is not {\em a priori } obvious if these are identical or not, in terms of either magnitude or temperature dependencies.

Surface mobility in glasses can be measured by the surface response to an external perturbation. In the past few decades, several forms of perturbations have been employed to characterize surface dynamics. A seminal method is provided by gold nanoparticles, in terms of both their embedding into the near-surface region~\cite{teichroeb2003direct, sharp2004properties, hutcheson2005nanosphere}, as well as surface flow around the nanoparticles \cite{qi2013molecular, daley2012comparing}. The response of a polymer surface to the introduction of a gold nanoparticle has been observed directly using transmission electron microscopy \cite{deshmukh2007direct}, and is similar to that observed for molecular glasses in Ref.~\cite{daley2012comparing}. Mainly, material was observed to accumulate towards the vicinity of the nanoparticle, before embedding of the latter eventually occurred. External surface perturbations can also be of the form of an initial surface morphology such as a nanohole \cite{fakhraai2008measuring}, or a nanostep \cite{chai2014direct}. More recently, Zhang {\em et al} \cite{zhang2016using, zhang2017invariant, zhang2017decoupling} have demonstrated that decorating the surface with tobacco mosaic virus (TMV) provides an alternative strategy. Another method to probe the surface mobility of glasses is to use nanobubbles that spontaneously nucleate on the glass surface when submerged into water~\cite{ren2020capillary}. An advantage of the nanosteps, the TMV decoration, and the nanobubbles is that these provide perturbations where the response of the surface can be described using  a two-dimensional surface-flow equation.

Recently, a quantitative study of surface mobility in stable glasses was performed for {\em N,N'}-bis(3-methylphenyl)-{\em N,N'}-diphenylbenzidine (TPD)  ($T_\text{g}$ = 330 K) \cite{zhang2017invariant}. In that work, the evolution of the surface of the glass in response to the presence of a TMV on the surface was measured. The measurements were performed on samples with a wide range (from 296 K to 330 K) of fictive temperatures $T_\text{f}$ characterizing the stability of the glass, and at two measurement temperatures of 296 K and 303 K. The surface diffusion constant was found to be independent of the stability of the glass as well as the origin of that stability (\textit{i.e.} aging from liquid cooled glass versus vapour deposition). While there were certainly differences in the surface mobility at the two temperatures considered, a more complete temperature dependence was not measured. In addition, the bulk dynamics were not measured but rather inferred from the Vogel-Fulcher-Tammann (VFT) time-temperature superposition, with parameters determined over a temperature region different from the window of $T_\text{f}$ studied.

In a separate study \cite{daley2012comparing}, the evolution of the surface of a rejuvenated 1,3-bis-(1-naphthyl)-5-(2-naphthyl)benzene (TNB, $T_\text{g}$ = 347 K) glass in response to gold nanoparticles placed on the surface was used to directly compare the temperature dependencies of surface and bulk relaxation processes in a molecular glass former. Surface flow exhibited a weak temperature dependence in that case, compared to bulk flow as characterized through embedding. Perhaps as expected, the temperature dependence of the time scale of nanoparticle embedding was well described there by bulk VFT parameters.

The surface mobility in glassy polystyrene (PS) has been studied  extensively \cite{tian2022surface,ediger2014dynamics} compared to most other glassy materials. The much larger molecular size of PS compared to the molecules used in Ref. \cite{zhang2017invariant} reopens the question as to whether liquid-cooled versus stable as-deposited glasses have the same surface properties, or not. In addition, there is an extension of the question to whether any form of stable PS glass is the same as the PS liquid-cooled glass, and how either of these compare to glassy PS films made from spin casting -- such as the relatively high $M_\text{w}$ samples of Ref.~\cite{fakhraai2008measuring}, and the low $M_\text{w}$ samples of Ref.~\cite{chai2014direct}.

\section{Experimental methods}
The original material is a broad distribution polystyrene ($M_\text{w} = 1200$ g/mol, catalogue no. 1024 from Scientific Polymer Products). The molecular weight distribution of the as-purchased material has been characterized in Ref.~\cite{raegen2020ultrastable}. The material was first thermally distilled at increasing temperatures into more monodisperse fractions in a vacuum chamber using a technique reported previously \cite{zhu2017evaporative}. The distilled product from 538 K was subsequently used as source material for the physical vapour deposition. Deposition was carried out in a Korvus Technologies HEX deposition unit with a base pressure of $10^{-5}$ mbar and a deposition rate of 0.05 nm/s \cite{raegen2020ultrastable}. The source material was heated in an ORCA temperature-controlled organic materials evaporation source. The source temperature used in generating the sample in this study was 514 K. The silicon substrate was attached to the sample stage which was cooled to a constant temperature of 283 K by a Peltier cooler. Ellipsometry measurements were performed with a J.A. Woollam M-2000 spectroscopic ellipsometer with a Linkam temperature-controlled stage. Heating and cooling rates for ellipsometric measurements were 10 K/min. The deposited film has a $T_\text{g}$ of 318 K and has a thickness of $\sim$ 100 nm as measured with ellipsometry, with a slight thickness gradient across the substrate. The average degree of polymerization ($\bar{N}$) of the sample is 9.02 and the polydispersity index (PDI) is 1.002, as determined with a Bruker Autoflex Speed MALDI-TOF (Matrix Assisted Laser Desorption/Ionization Time of Flight) mass spectrometer.

Aqueous solutions of gold nanoparticles with diameter $\sim$ 20 nm were produced by the standard citrate reduction technique \cite{turkevich1951study}. For as-deposited films, the gold nanoparticle solution was directly dropped onto the surface, and after 1-2 min the film was tilted to allow the  droplet to flow to the side of the Si wafer and onto a wipe for removal. To study rejuvenated glasses, the as-deposited films were heated to 343 K ($T_\text{g} + 25$ K) for 3 min and cooled in ambient air to room temperature before introducing the nanoparticles on the surface. This thermal treatment was found to be sufficient to produce ordinary liquid-cooled glasses. The nanoparticles on the glassy PS surface were imaged using a JPK Nanowizard 3 atomic force microscope (AFM) operating in tapping mode. The surface coverage density was $\sim$ 5 particles/100 ${\mu}$m$^2$ based on AFM measurements. For each film studied at a specific temperature, two nanoparticles were chosen that had  at least 500 nm separation from any other nanoparticle (to minimize effects from more than one particle). Once a pair of such particles were located, the AFM hot stage (JPK High Temperature Heating Stage) was set to a certain temperature, and 2 $\mu$m $\times$ 2 $\mu$m regions centered on each nanoparticle were scanned at different times. 

\section{Theoretical modelling}
\begin{figure*}
\centering
\includegraphics[width=0.5\textwidth]{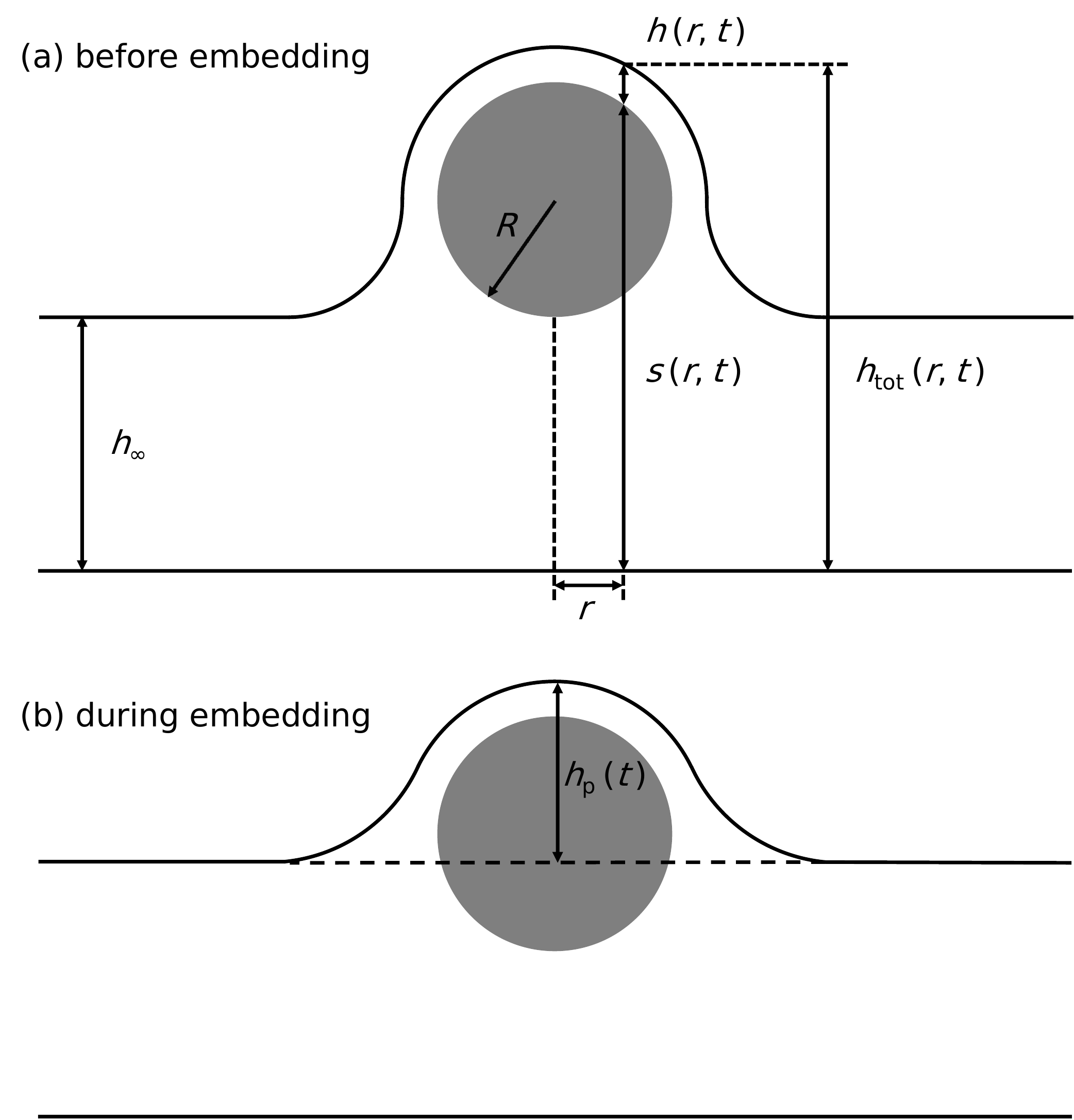}
\caption{Schematic diagram of the relevant processes discussed in the text. Variables used in the calculations and discussed throughout the text are shown in the schematic.}
\label{fig1}
\end{figure*}

We consider a flat glassy polymer film of thickness $h_{\infty}$ placed on the surface of a horizontal and flat rigid substrate located at vertical coordinate $z = 0$. A nanoparticle of radius $R$ is placed on the surface of the film. We assume a favourable wetting condition and the presence of a liquid-like mobile layer of thickness $h^*$ (with $h^* \ll h_{\infty}$) at the free surface of the glass. The system then evolves through a three-step process \cite{deshmukh2007direct}: i) at short time, some surface polymeric material rapidly migrates and fully coats the immobile particle; ii) then, the glass-air interface evolves by surface flow across the still-immobile particle; iii) at long times, the particle eventually embeds in the film, under the action of the capillary pressure. In this last step, the dynamics is limited by the bulk viscous Stokes-like drag acting on the particle. Fig.~\ref{fig1} provides a schematic of the processes involved for surface flow and eventual embedding.

We focus only on the second step for now, as it is well separated in time from the two other steps. By considering the very large viscosity of the bulk glass, and that there is only a thin film of flowing material, the interface dynamics can be described by adopting the lubrication theory \cite{BatchelorBook, oron1997long}. The flow is assumed to be incompressible and axisymmetric. Since the flow is localized only near the free surface, the particle can be modelled as an effective substrate, described by its profile $s(r) = \Theta(R - r)[h_{\infty} + R + \sqrt{R^2  - r^2}]$, with $r$ the radial coordinate and where $\Theta$ is the Heaviside function. The glass-air interface profile is defined as $h_{\textrm{tot}}(r, t) = h(r, t) + s(r)$, where $h(r,t)$ is the glass thickness profile. We define the initial profile $h_{\textrm{tot}}(r,0)$, from the earliest experimental data available. At time $t=0$, we assume the particle to be already covered by a nanometric layer of polymeric material.

The hydrodynamic flow is described by the pressure field $p(r,z,t)$, in excess to the atmospheric pressure, and the velocity field $\mathbf{v}(r,z,t)$. The material parameters are the polymer-air surface tension $\gamma$, and the viscosity of the surface layer $\eta$, both assumed to be constant at a given temperature. At the free surface, we impose no shear, \textit{i.e.} $\partial_z(\mathbf{v}\cdot\mathbf{e}_r)|_{z=h_{\textrm{tot}}} = 0$, where $\partial_i\equiv \partial /\partial i$, and where $\mathbf{e}_r$ is the unit vector in the radial direction. We also impose no slip at the bottom of the mobile layer, \textit{i.e.} $\mathbf{v}\cdot\mathbf{e}_r |_{z=h_{\textrm{tot}} - h^*} = 0$.

Within the lubrication theory introduced above, the flow is mainly horizontal, \textit{i.e.} $\mathbf{v}=v\, \mathbf{e}_r$. Invoking the Stokes equation, the pressure $p$ is thus invariant in the vertical direction, and by integration the velocity profile in the radial direction is found to be:
\begin{equation}
v(r,z,t)= \frac{1}{2\eta}\left(z^2 + h_{\textrm{tot}}^2 - h^{*2} - 2zh_{\textrm{tot}}\right)\partial_r p\ ,
\end{equation}
which corresponds to the familiar Poiseuille flow for $h_{\textrm{tot}} - h^* \leq z\leq h_{\textrm{tot}} $. For $0 \leq z < h_{\textrm{tot}} - h^*$, we assume zero velocity, since the bulk glass is immobile to lowest order in the description. Furthermore, volume conservation requires that:
\begin{equation}
\partial_t h + \frac{1}{r}\partial_r\left(r\int_{h_{\textrm{tot}}-h^*}^{h_{\textrm{tot}}}\textrm{d}z\, v\right)  = 0\ .
\end{equation}
The excess pressure $p(r,t)$ can be evaluated at the free interface, and contains two terms: a capillary contribution from the Young-Laplace equation, and an ad-hoc repulsive disjoining contribution~\cite{eggers2005contact,poulain2022elastohydrodynamics} ensuring the wettability condition and thus preventing film rupture at the particle tip. All together, and in the limit of small interface slopes, one gets:
\begin{equation}
p  = -\gamma r^{-1}\partial_r(r\partial_r h_{\textrm{tot}}) + B(h_{\textrm{eq}}/h)^9\ ,
\end{equation}
where $B$ sets the magnitude of the disjoining pressure, and $h_{\textrm{eq}}$ is the equilibrium film height. By combining the equations above, we derive the Glassy Thin Film Equation \textit{atop a Nanoparticle} (GTFEN):
\begin{equation}
\partial_t h(r,t) + \frac{\gamma h^{*3}}{3\eta r}\partial_r\left\{r\partial_r\left( r^{-1}\partial_r\left[ r\partial_r h(r,t) + r\partial_r s(r)\right] - Bh_{\textrm{eq}}^9/[\gamma h(r,t)^9] \right) \right\} = 0\ .
\label{GTFEN}
\end{equation}
Written this way, we see that the nanoparticle, through the effective substrate profile $s(r)$ and the disjoining pressure, generates a forcing source within the free Glassy Thin Film Equation (GTFE)~\cite{chai2014direct}.

In contrast to simpler and linear versions of GTFE~\cite{salez2012capillary,pedersen2021nanobubble}, the complex effective substrate and the nonlinear disjoining term in GTFEN require us to perform a numerical integration. We thus introduce dimensionless variables, through: $X=r/R$, $\mathcal{T}=t\gamma h^{*3}/(3\eta R^4)$, $S(X)=s(r)/R$, $H(X,\mathcal{T})=h(r,t)/R$ and $P(X,\mathcal{T})=Rp(r,t)/\gamma$. We also fix the two dimensionless parameters of the problem, as follows: i) the aspect ratio $h_{\infty}/R$ is directly estimated from the experimental $h_{\infty}$ and $R$ values, for each given sample, which in turn fully determines $S(X)$; ii) the dimensionless disjoining magnitude is arbitrarily fixed to $Bh_{\textrm{eq}}^9/(\gamma R^8)=10^{-9}$, which for physical parameters relevant to PS on gold in air corresponds to a nanometric $h_{\textrm{eq}}$. Note however that the late-time interface dynamics is insensitive to small variations around the chosen disjoining magnitude. The dimensionless form of Eq.~\eqref{GTFEN} is solved numerically by using a finite-element scheme. The initial condition $H(X,0)=h_{\textrm{tot}}(R.X,0)/R-S(X)$ is directly constructed from the experimental free surface profile $h_{\textrm{tot}}(r,0)$ measured at $t=0$. At all times $\mathcal{T}>0$, we also impose four boundary conditions, namely the vanishing of both $\partial_XH$ and $\partial_XP$, at both $X=0$ and $X\rightarrow \infty$ (\textit{i.e.} the far-field bound of the numerical spatial domain).

From fitting the obtained numerical profiles to the experimental ones, and invoking the PS-air surface tension $\gamma \approx 40$~mN/m, we can thus now measure the surface mobility $M = h^{*3}/(3\eta)$ as a single free parameter, for each given sample type and temperature.

\section{Results}
\begin{figure*}
\centering
\includegraphics[width=0.5\textwidth]{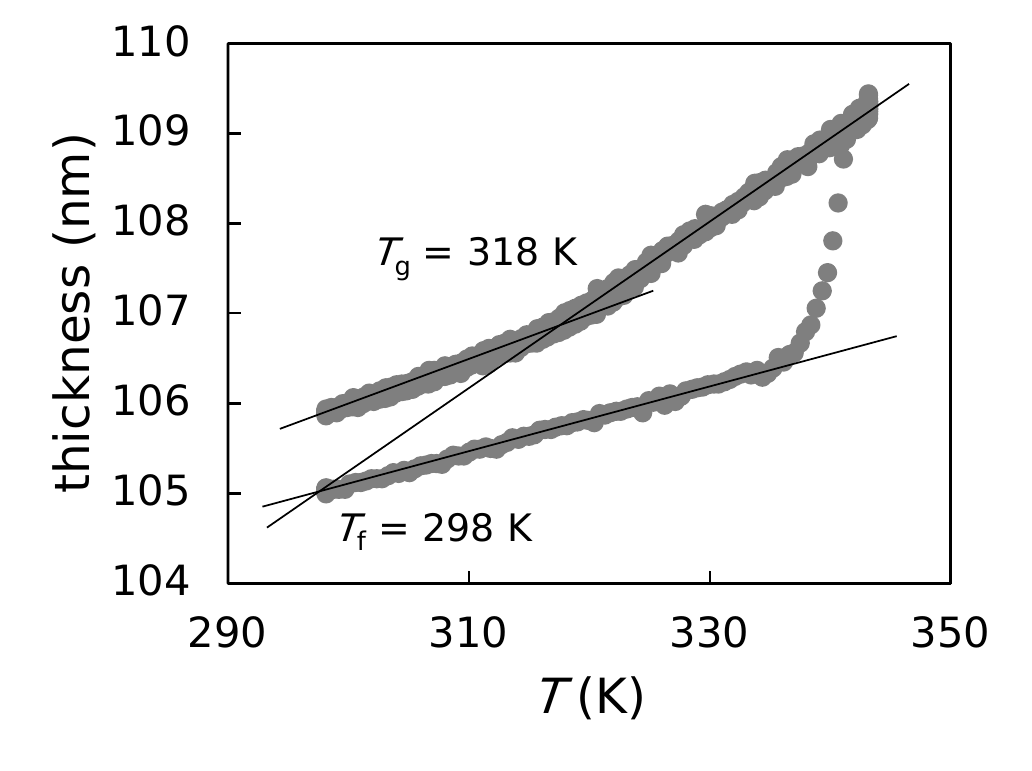}
\caption{Total film thickness versus temperature obtained from ellipsometric scans of the as-deposited samples. The heating (bottom) and subsequent cooling (top) branches are performed with equal rates of 10 K/min. On the first heating the film is held at the highest temperature (343 K) for 2.5 min to ensure full rejuvenation from the stable glass to the normal supercooled liquid, before cooling in the supercooled liquid state and down to the rejuvenated glass. These scans are used to determine the fictive temperature $T_\text{f}$, defined from the intersection between the stable glass and supercooled liquid lines, as well as the glass transition temperature $T_\text{g}$, defined from the intersection between the supercooled liquid and rejuvenated glass lines.}
\label{fig2}
\end{figure*}

First, Fig.~\ref{fig2} shows the ellipsometrically determined film thickness versus temperature $T$, for an as-deposited supported PS film without nanoparticles. The data in these scans are used to determine the fictive temperature $T_\text{f}=298$ K, as well as the glass transition temperature $T_\text{g}=318$ K. Comparing the thickness of the as-deposited sample versus the one after a full heat/cool cycle is used to determine the increased density of the stable glass. In this case the increase in density of stable glass versus normal glass is $\sim$ 1\% The data also demonstrates enhanced kinetic stability since the material apparently remains in the glassy state even after  the temperature has been raised above $T_\text{g}$. By comparing the $T_\text{f}$ and $T_\text{g}$ with the ones in Ref.~\cite{raegen2020ultrastable}, we can estimate that the as-deposited material in the present study is analogous to a liquid-cooled glass that would have been aged at $T_\text{f}$ for $\sim300$ years. It is worth noting that this estimate is based on an Arrhenius law, while using the more conventional VFT law would result in orders of magnitude larger estimates for aging times.

\begin{figure*}
\centering
\includegraphics[width=0.8\textwidth]{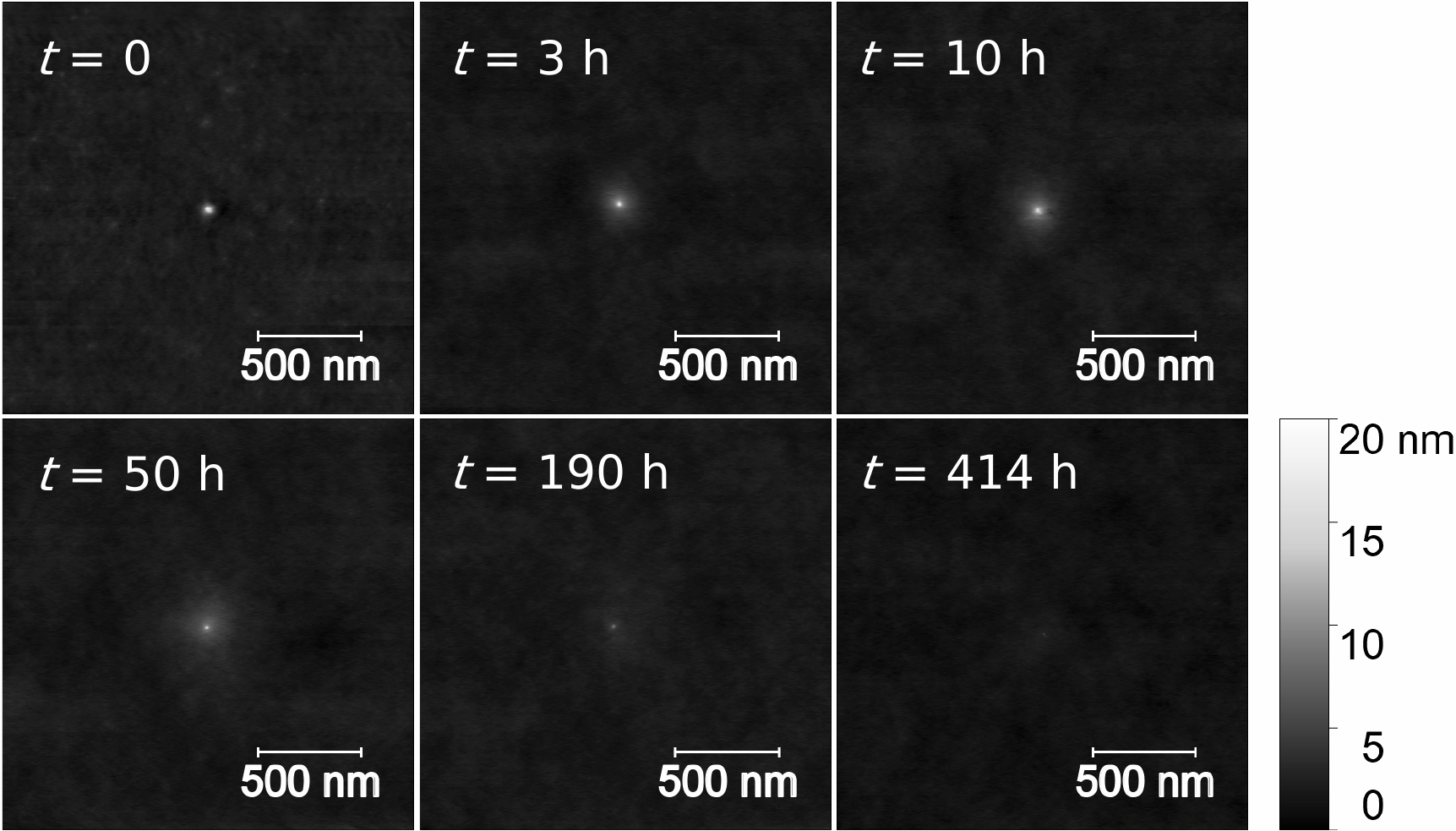}
\caption{AFM images showing the build up and levelling of polymeric material around a gold nanoparticle, as well as the subsequent nanoparticle embedding, for a rejuvenated glassy polystyrene film at $T_\text{g}-5$ K (313 K).}
\label{fig3}
\end{figure*}

Now, we consider the surface response of the glassy films after nanoparticles have been placed on the surface. Fig.~\ref{fig3} shows AFM images of a 16 nm nanoparticle on a rejuvenated PS film at a temperature of 313 K, \textit{i.e.} 5 K below $T_\text{g}$. In the $t=0$ panel, we see a sharp image of the nanoparticle surrounded by an approximately circular dark region. The latter is a depletion zone where material has moved from in order to cover the nanoparticle.  As the evolution progresses from 0 h to 10 h, we can see that there is a brighter area immediately surrounding the nanoparticle, and the depletion zone has moved outwards. By 50 h of evolution, in addition to a continual buildup of material, we can see that the apparent height of the nanoparticle is also decreasing. This indicates an embedding or engulfment of material which can only occur when there is sufficient bulk mobility (\textit{i.e.} the material is above or very near its $T_\text{g}$). After 414 h of evolution, the nanoparticle has almost completely embedded into the underlying polymer.

\begin{figure*}
\centering
\includegraphics[width=0.95\textwidth]{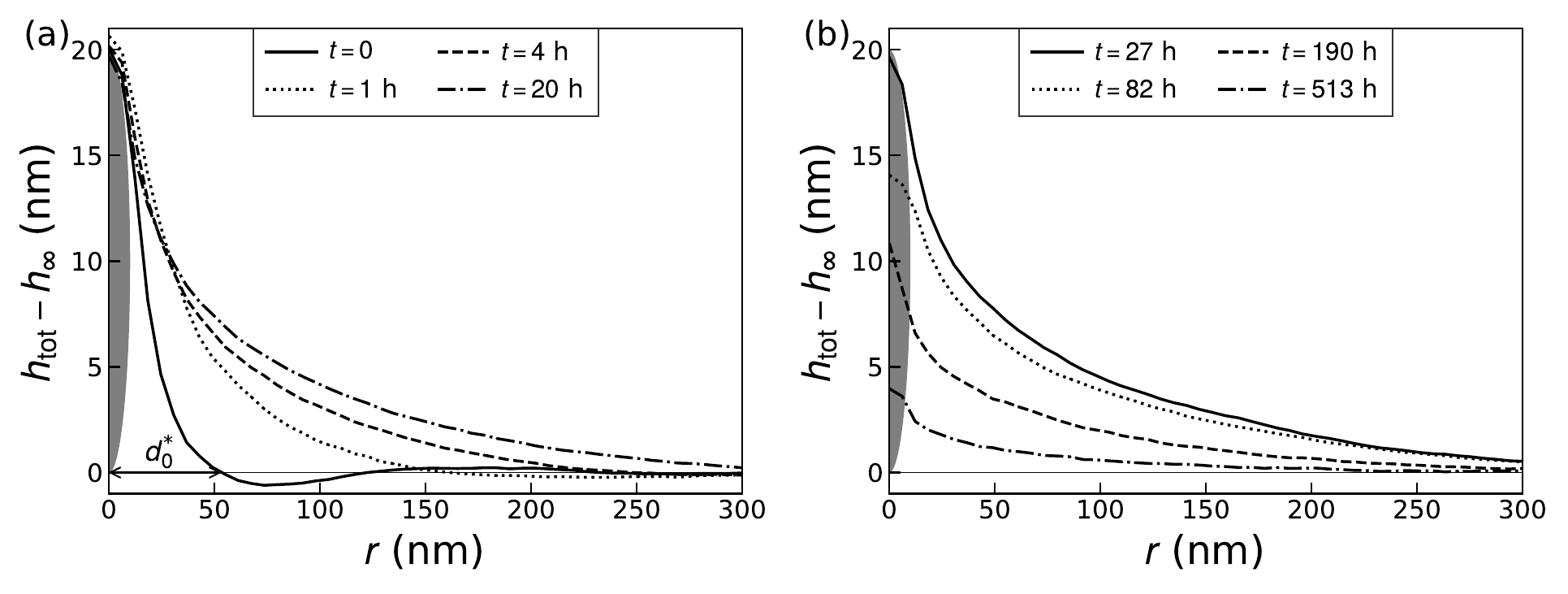}
\caption{Surface profiles at various times as indicated in legend, of a rejuvenated film at $T_\text{g}-5$ K (313 K). (a) The short-term evolution before embedding occurs. (b) The long-term embedding process. The grey shaded areas indicate the shape of the spherical nanoparticle with a diameter of 20 nm, at $t=0$. The solid horizontal line indicates the base line reference. $d^*_\text{0}$ shown in (a) is the width $d^*(t)$ (defined in text) at $t=0$.}
\label{fig4}
\end{figure*}

Fig.~\ref{fig4} shows the radial profiles of the PS-air interface along time, for another nanoparticle but on the same sample and at the same temperature as the AFM images in Fig.~\ref{fig3}. Each profile $h_{\textrm{tot}}(r,t)$ was obtained by angularly averaging the image from the centre of the nanoparticle, and subtracting the base-line height. To determine the latter, a mask was applied to the AFM image, covering the nanoparticle and the depletion zone, and the average height of the 2 $\mu$m $\times$ 2 $\mu$m scan area excluding the mask area was measured. The dark solid line in Fig.~\ref{fig4}(a) represents the profile at $t=0$. In this initial profile, we can see a well defined depletion zone with a maximum depletion at about 70 nm from the centre of the nanoparticle. At larger times, there is more and more material accumulation near the nanoparticle, while the depletion zone becomes less pronounced and moves further out from the nanoparticle's vicinity. Note that the depletion zone was clearly evident in the AFM images in Fig.~\ref{fig3}, and was also observed in the surface evolution of the TNB molecular glass in similar conditions \cite{daley2012comparing}. From the data in Fig.~\ref{fig4}, we can extract two parameters that characterize separately the near surface and bulk dynamics. For the bulk dynamics, the natural observable is the height $h_{\textrm{p}}(t)$ of the nanoparticle. At temperatures near $T_\text{g}$, and for sufficiently long times (such as the ones in Fig.~\ref{fig4}(b)), the bulk of the material has a large-enough mobility, so that the nanoparticle can embed, and $h_{\textrm{p}}$ will evolve over time. For the surface dynamics, we define a typical horizontal width $d^*(t)$ of the profile as the minimal value of the radial coordinate $r$ at which the height crosses zero (\textit{i.e.} the base line).

\begin{figure*}
\centering
\includegraphics[width=0.5\textwidth]{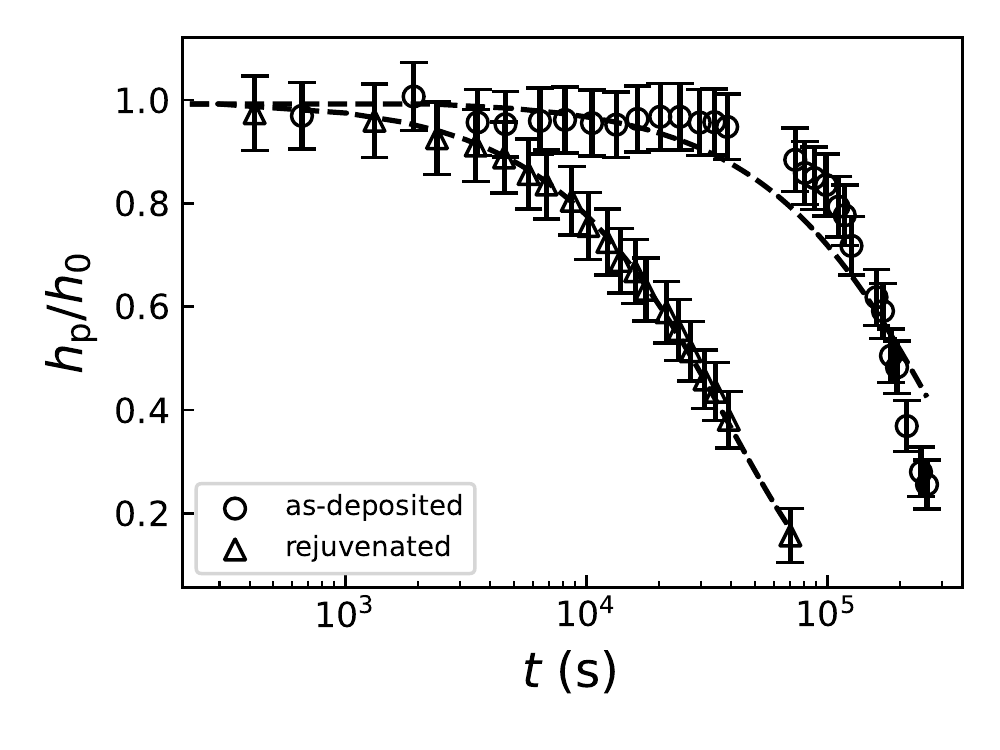}
\caption{$h_{\textrm{p}}/h_0$ as a function of time $t$, in lin-log representation, where $h_0\equiv h_{\textrm{p}}(0)$ is the initial particle height, for both the as-deposited (circles) and rejuvenated (triangles) films, at $T_\text{g}$ (318 K). The dashed lines are best exponential fits.}
\label{fig5}
\end{figure*}

Focusing first on the bulk dynamics, Fig.~\ref{fig5} shows the normalized particle height, $h_{\textrm{p}}/h_0$ as a function of time for both the as-deposited and rejuvenated films at $T=T_\text{g}$. There are two notable differences between the embedding behaviours in the two systems. The most obvious difference is that, although particle embedding occurs in both the as-deposited and rejuvenated films at this temperature, the embedding occurs earlier in the rejuvenated film. This is true in all cases where embedding can be observed for both types of film. Embedding occurs in the as-deposited films at about an order of magnitude longer time than for the rejuvenated films, in all such cases. Another difference is the shape of the embedding curve. For the rejuvenated (\textit{i.e.} normal glass) film, the exponential fit is an excellent characterization of the data, while for the as-deposited (\textit{i.e.} stable glass) film, there is a significant deviation from an exponential behaviour. In fact, for the as-deposited film, the embedding is better characterized by an affine relation between the apparent height and $\log{t}$ for times larger than the embedding start time. Unlike the early build-up of material near the nanoparticle, which can occur through surface mobility only, complete embedding requires a bulk response of the polymer. Therefore, the embedding dynamics of the rejuvenated sample would  correspond to the bulk response of the material at $T_\text{g}=318$~K, while the embedding dynamics of the as-deposited sample would correspond to the bulk response of the material at $T_{\textrm{f}}=298$~K, by definition of the fictive temperature (see Fig.~\ref{fig2}). Interestingly, with these temperatures, we would have expected the ratio between the two embedding times to be much greater (\textit{i.e.} $\sim$7 decades in time). We can rationalize the discrepancy between this expectation and the observation by comparing to the rejuvenation times from Ref.~\cite{raegen2020ultrastable}. In that case, we can see that for $T_\text{g}/T$ = 318 K/318 K = 1, the rejuvenation time (\textit{i.e.} the time for the material to fully convert from a stable glass to a normal supercooled liquid) is about $10^5$~s. The latter value compares well with the embedding time of the as-despited sample in Fig.~\ref{fig5}. This coincidence suggests that the unexpectedly fast embedding in the as-deposited sample is due to the fact that the sample is actually rejuvenating during the measurements, and the nanoparticle is embedding into the rejuvenated material. This is also consistent with the peculiar shape of the embedding curve for the as-deposited sample, since an affine law is a typical functional form for a rejuvenation front travelling through the film.

\begin{figure*}
\centering
\includegraphics[width=0.5\textwidth]{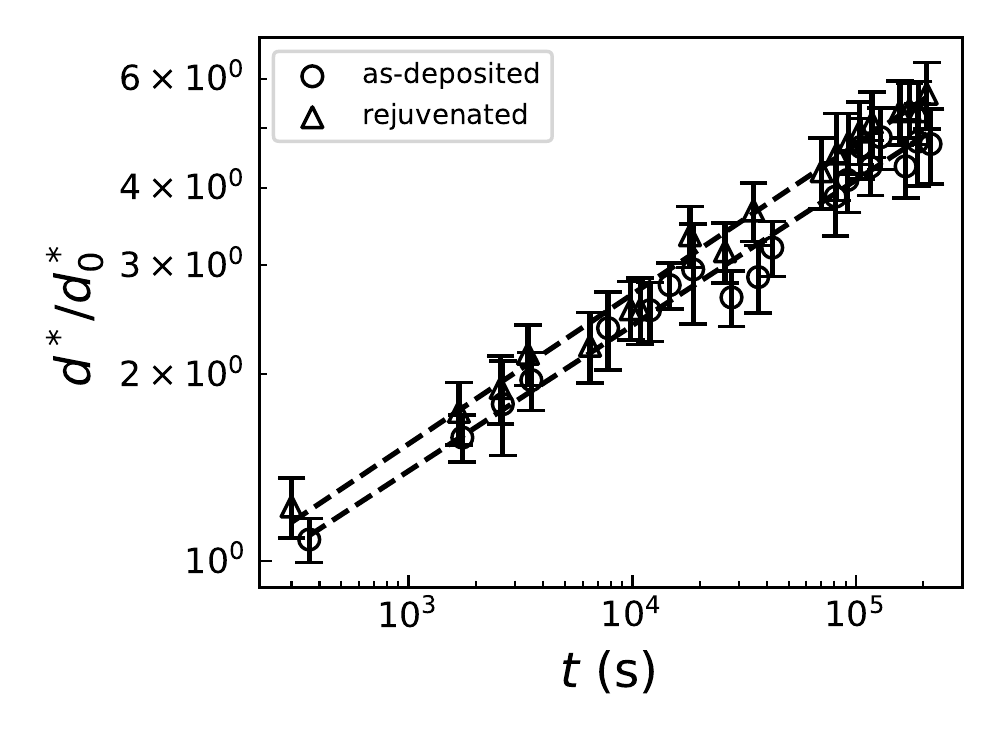}
\caption{Normalized width as a function of time, in log-log representation, where $d^*_0\equiv d^*(0)$ is the initial width, for both the as-deposited (circles) and rejuvenated (triangles) samples, at $T_\text{g}-10$ K (308 K). The dashed lines are best power-law fits, with exponents $0.23\pm0.02$ (bottom) and $0.24\pm0.02$ (top).}
\label{fig6}
\end{figure*}

Turning now to the  the surface dynamics, Fig.~\ref{fig6} displays the normalized width $d^*/ d^*_0$ as a function of time, for both the as-deposited and rejuvenated samples at $T_\text{g}-10$ K. It is clear from this figure that, both qualitatively and quantitatively, there is no discernible difference between the surface flows of as-deposited and rejuvenated glasses. Both types of glass exhibit a mobility enhancement at the surface as compared to the bulk -- a feature that has been widely reported for many different glass formers \cite{teichroeb2003direct, fakhraai2008measuring, chai2014direct, ren2020capillary, zhu2011surface, malshe2011evolution, cao2015high}. Moreover, the surface mobilities of the two types of glass studied here are quantitatively the same, within our error bars. Similar plots (not shown) for the entire collection of data at 7 different temperatures indicate that, while there is certainly some scatter in the data,  the enhanced surface flows of stable PS glasses and normal PS glasses do not exhibit  any discernible difference. Furthermore, Fig.~\ref{fig6} indicates the existence of a power law between $d^*$ and $t$. It has been shown previously~\cite{chai2014direct}, that, if the dynamics is governed by capillary-driven surface viscous flow, the associated power-law exponent should be equal to 1/4. The best fits in Fig.~\ref{fig6} give exponent values of $0.23 \pm 0.02$ for the as-deposited sample, and $0.24 \pm 0.02$ for the rejuvenated sample. Similar best fits (not shown) for the entire collection of data yield exponents in the $0.17-0.25$ range. Data at the highest temperatures (where the surface relaxation is more advanced) exhibit exponents closer to $1/4$, as compared to data at lower temperatures. This is similar to the results in Ref.~\cite{chai2020using}, where more-evolved profiles were comparatively better described by the expected asymptotic exponent. One might be concerned that the reason the surfaces of as-deposited and rejuvenated films are so similar is that the as-deposited films have at least partially rejuvenated.  However, the data of Fig.~\ref{fig5} shows that for $T=318$ K rejuvenation does not begin to occur until $t \sim 5 \times 10^4$ s.  For lower temperatures, that onset will be at even greater times \cite{raegen2020ultrastable}. This comparison provides confidence that the surface evolution of as-deposited films is not being influenced by rejuvenation.

\begin{figure*}
\centering
\includegraphics[width=0.95\textwidth]{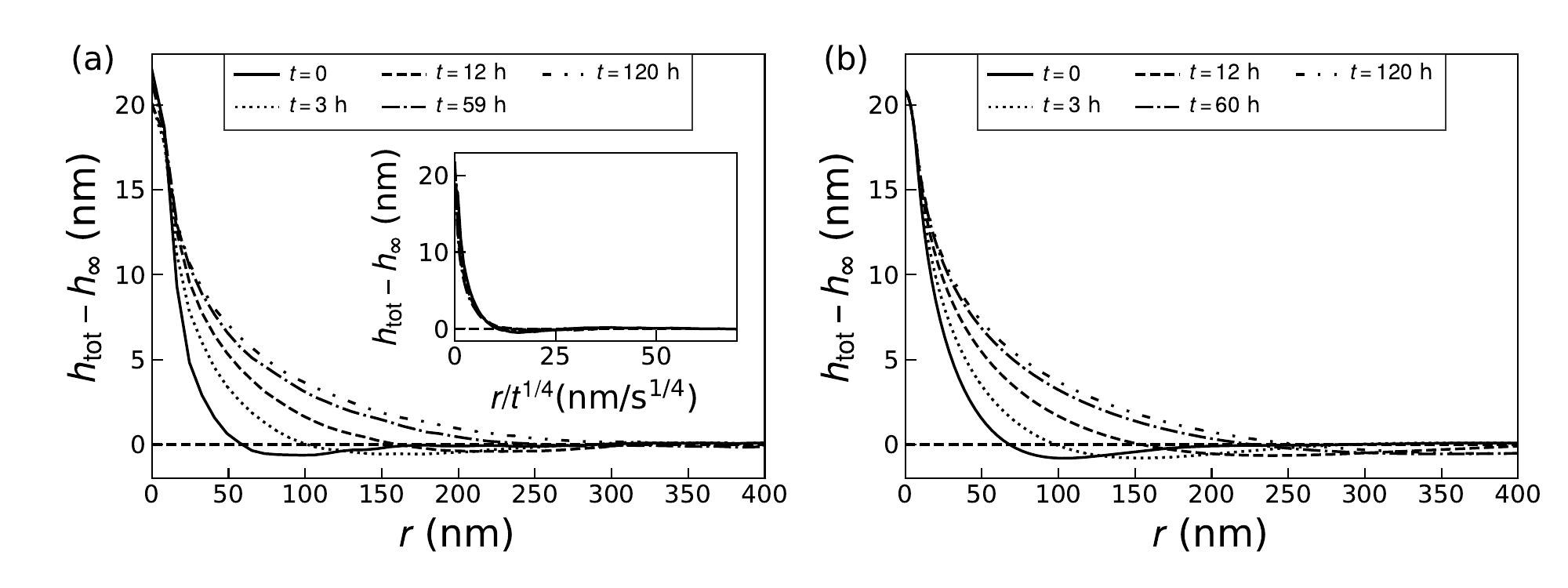}
\caption{(a) Experimental surface profiles at various times as indicated in legend, of an as-deposited film at $T_\text{g}-15$ K (303 K). The horizontal dashed line indicates the base line reference. (Inset) Same data, but with the indicated rescaling of the $x$-axis. (b) Numerical solutions of Eq.~\eqref{GTFEN} at various times, as indicated, using the experimental initial profile of panel (a) as an input and the surface mobility as a single fit parameter.}
\label{fig7}
\end{figure*}

Fig.~\ref{fig7}(a) shows surface profiles for an as-deposited film at $T_\text{g}-15$~K, for several times where embedding has not yet occurred. Previous works~\cite{chai2014direct,zhang2017invariant} and the fact that we observe $d^*\propto t^{1/4}$ suggest that the data is self-similar, and that replacing $r$ on the $x$-axis by $\frac{r}{t^{1/4}}$ should result in a collapse of all the curves. The inset of Fig.~\ref{fig7}(a) shows that this is indeed the case. To go beyond scaling, and quantitatively check the validity of the developed GTFEN model based on capillary-driven surface viscous flow, we fit the experimental profiles by the numerical solutions of Eq.~\eqref{GTFEN}. The observed agreement between Fig.~\ref{fig7}(a) and Fig.~\ref{fig7}(b) confirms the validity of the GTFEN.

\begin{figure*}
\centering
\includegraphics[width=0.95\textwidth]{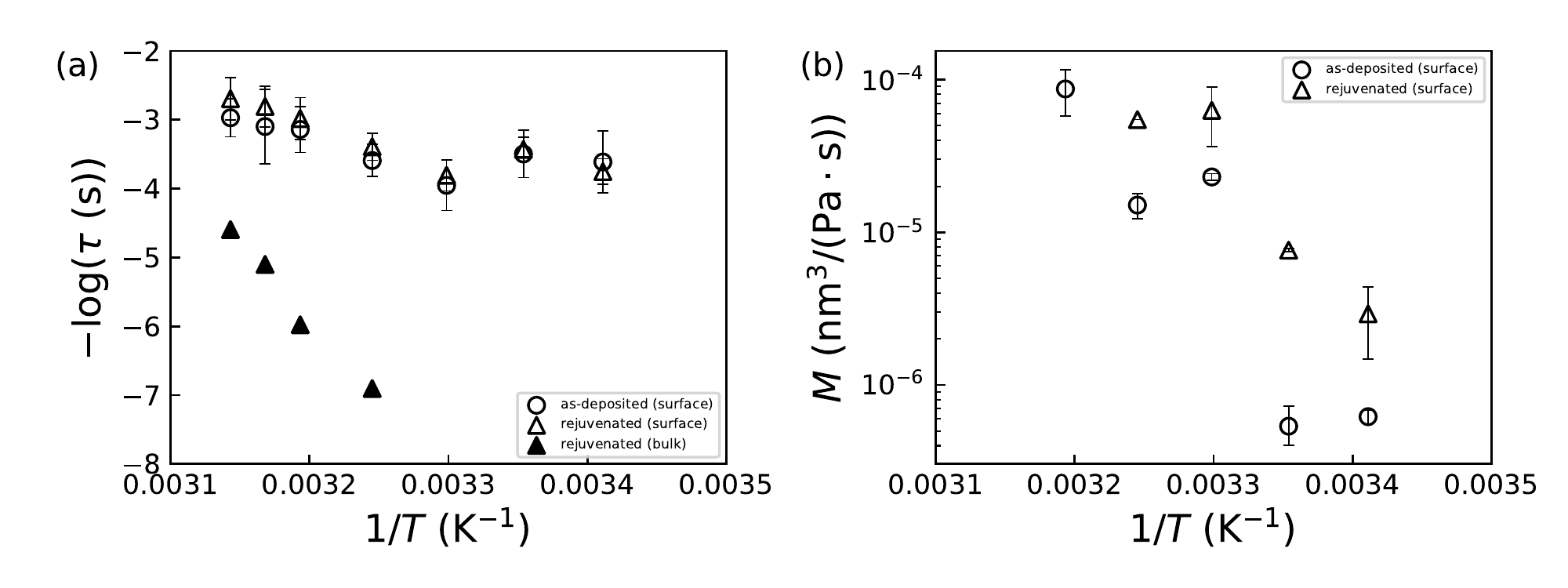}
\caption{(a) Inverse relaxation time $1/\tau$ as a function of the inverse temperature $1/T$, in log-lin representation, as extracted from either the best fit to an exponential embedding law (bulk, see Fig.~\ref{fig4}) or the width-doubling time (surface, see Fig.~\ref{fig6}), for two types of samples, as indicated in legend. (b) Surface mobility $M$ as a function of the inverse temperature $1/T$, in log-lin representation, as extracted from the best fit of the experimental surface profile to the numerical solution of Eq.~\eqref{GTFEN} (see Fig.~\ref{fig7}(b)), for the same two types of samples, as indicated in legend.}
\label{fig8}
\end{figure*}

We now aim at quantifying the temperature dependencies of the bulk and surface processes. The relaxation time $\tau$ of the bulk is obtained from the best exponential fit $\exp(-t/\tau)$ of the $h_{\textrm{p}}(t)/h_0$ data in Fig.~\ref{fig5}. The relaxation time $\tau$ of the surface is empirically defined as the time needed for the width $d^*$ to increase from $d^*_0$ to $2d^*_0$ (see Fig.~\ref{fig6}). We note that the choice of $d^*$ is somewhat arbitrary.  For example in  ~\cite{zhang2017invariant},  the width of the surface profile at a height of 2 nm was monitored. Fig.~\ref{fig8}(a) displays the temperature dependencies of both these times, for both the as-deposited and rejuvenated samples. For bulk relaxation, despite the difference in magnitudes already discussed above, we observe that characteristic times derived from the as-deposited and rejuvenated data display very similar temperature dependencies. This might be expected, since the as-deposited sample is rejuvenating during the embedding (as discussed above). For surface relaxation times, one observes two qualitative differences with respect to the bulk trend. First, surface relaxation occurs on a much shorter time scale. Secondly,  surface relaxation times have a much weaker temperature dependence than the bulk one, which even seems to vanish at low temperature. Nevertheless, the difference between the bulk and surface relaxation times decreases as the temperature is increased. Extrapolations of the bulk and surface trends even suggest that the two processes would have the same relaxation time for $T_\text{g}/T=0.98$. The behaviour exhibited in Fig.~\ref{fig8}(a) is similar to that reported for spin-coated PS~\cite{fakhraai2008measuring}, as well as for vapour-deposited TNB~\cite{daley2012comparing}. 

In Fig.~\ref{fig8}(b), we show the temperature dependence of the surface mobility $M$, as obtained from the best fits of the experimental surface profiles to the numerical solutions of Eq.~\eqref{GTFEN} (see Fig.~\ref{fig7}(b)). Here, one observes a continuous, Arrhenius-like decay of the mobility as the temperature is lowered, which is qualitatively similar to the behaviour reported for spin-coated PS~\cite{chai2014direct}. However, the surface mobility values in Ref.~\cite{chai2014direct} are comparatively smaller, and in some cases by even a few orders of magnitude. We can understand this observation by taking into account the $M_\text{w}$ dependence of the mobility. Indeed, in Ref.~\cite{chai2020using}, it was shown that for $T<T_\text{g}$ the surface mobility can vary by as much as four orders of magnitude for a one order of magnitude change in $M_\text{w}$. It is difficult to be more quantitative, because the samples used in Ref.~\cite{chai2020using} were much more polydisperse than the samples considered in this work -- due to the intrinsic mass-selective nature of vapour deposition -- and it is not clear what the effect of polydispersity is on the mobility. We can compare these mobilities to the relaxation times from Fig.~\ref{fig8}(a).  Because the mobility values are given by $(h^*)^3/(3 \eta)$ we would might expect that conclusions determined from the temperature dependence of the measured relaxation times would be the same as those arising from the mobility. Differences between Fig.~\ref{fig8}(a) and Fig.~\ref{fig8}(b) show that this is surprisingly not the case. We have investigated these differences in detail in the Supplementary Information. The temperature dependence of the mobilities is the same within error for the as-deposited and rejuvenated films, but the magnitudes of the mobilities of the as-deposited materials are slightly lower than those of the rejuvenated materials. 

\section{Discussion}
\begin{figure*}
\centering
\includegraphics[width=0.5\textwidth]{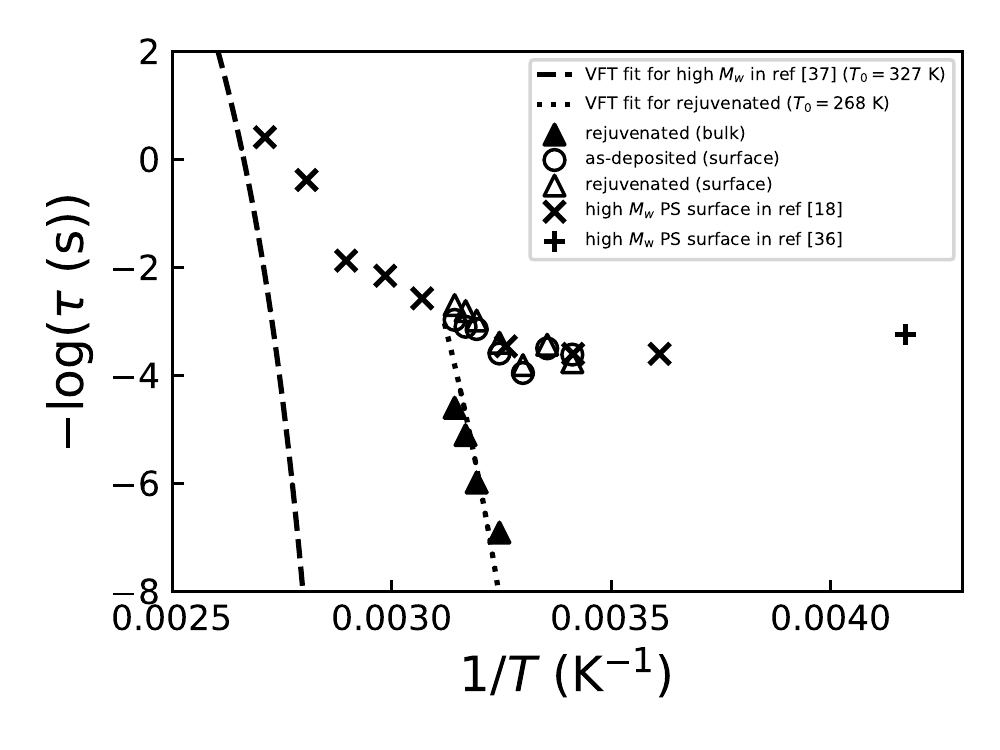}
\caption{ Inverse relaxation time $1/\tau$ as a function of the inverse temperature $1/T$, in log-lin representation, as extracted from either the best fit to an exponential embedding law (bulk, see Fig.~\ref{fig4}) or the width-doubling time (surface, see Fig.~\ref{fig5}), for two types of samples, as indicated in legend. For comparison, are also shown the high-$M_{\textrm{w}}$ data of Ref.~\cite{fakhraai2008measuring} and Ref.~\cite{qi2009near}. The dashed line represents the VFT behaviour of high-$M_{\textrm{w}}$ bulk PS~\cite{dhinojwala1994rotational}, while the dotted line represents the VFT behaviour of low-$M_{\textrm{w}}$ bulk PS.}
\label{fig9}
\end{figure*}

The fact that the surfaces of the as-deposited and rejuvenated glasses have such similar dynamics is a significant observation. Indeed, it is not necessarily expected that the surface mobility of a material assembled using vapour deposition should be the same as the one of a sample prepared by cooling a liquid. A similar observation was first reported for molecular glasses~\cite{zhang2017invariant}, where the values of the surface diffusion constants for as-deposited, aged, and rejuvenated materials were found to be indistinguishable, for each of the two temperatures considered therein. Our work provides a complementary approach, where we have considered as-deposited and rejuvenated glasses at a number of different temperatures, allowing for the temperature dependencies to be investigated in details. Note that, in Ref.~\cite{zhang2017invariant}, the $T_\text{f}$ values ranged from 296 K to 330 K, while the $T_\text{g}$ value was 330 K, which gives a range of $T_\text{g}-T_\text{f}$ of $0-34$~K.  The $T_\text{g}$ of the material in the current work is 318 K, and the $T_\text{f}$ of the stable glass is 298 K, which gives $T_\text{g}-T_\text{f} = 20$ K. This is close to the middle of the range of Ref.~\cite{zhang2017invariant}, thus allowing for a reasonable comparison.

Additionally, glassy PS is a system whose surface has been extensively studied in the literature, for $M_\text{w}$ values varying by about three orders of magnitude, encompassing oligomeric materials as well as highly entangled systems. We can therefore easily compare our low-$M_{\textrm{w}}$ results to others. In this context, two particularly relevant studies are the ones for high $M_{\textrm{w}}$ reported in Refs.~\cite{fakhraai2008measuring,qi2009near}. In Fig.~\ref{fig9}, we thus compare our data on the temperature dependencies of the bulk and surface relaxation times, to the ones from Refs.~\cite{fakhraai2008measuring,qi2009near}. Strikingly, for the surface data, there seems to be a universal change from an Arrhenius-like behaviour to an athermal one, as the temperature is decreased. From the Arrhenius-like region, we can estimate a typical activation barrier of $\sim 10^4$~K for both systems. The crossover seems to happen at $1/T \sim 3.25 \times 10^{-3}\ \text{K}^{-1}$, independently of the $T_\text{g}$ values of the materials (which differ by about 60 K). The actual relaxation times from the current study as well as ~\cite{fakhraai2008measuring,qi2009near} agree with each other quantitatively, but this is likely coincidental as the probed physical phenomena were different in the two cases. Nevertheless, given that the materials studied in the current work are nearly oligomeric, since $M_\text{w} <1000$ and because it was prepared by vapour deposition, and given that the materials in Ref.~\cite{fakhraai2008measuring} have a $M_\text{w}$ value of 640,000 g/mol and were prepared by spincoating out of solvent, this apparent coincidence is truly remarkable. It may in fact indicate that the dynamical properties of the mobile surface layers on glassy films are somewhat universal for all PS materials. 

The comparison made in the previous paragraph can be qualitatively extended to the measurements of surface mobility in ultrastable metallic glasses \cite{luo2018ultrastable}. Indeed, the similarity between the temperature dependencies of  the surface mobilities of PS, from the oligomeric materials used here and in the literature~\cite{chai2014direct}, to the large-$M_{\textrm{w}}$ materials studied previously~\cite{chai2020using}, and the temperature dependence of surface mobility in metallic glasses is striking too. As discussed above, it is interesting to note the absence of any saturation in the surface mobility at low temperature (see Fig.~\ref{fig8}(b)), in contrast to the observation made for the surface relaxation time (see Figs.~\ref{fig8}(a) and~\ref{fig9}). This subtle apparent discrepancy may simply result from the difference in the considered physical observables, but it may also highlight deeper differences at low temperatures between the quantitative modelling and the simple empirical criterion (see Supplementary Information).

Finally, the current work may have some relevance for the continuing discussion around the anomalous glass transition temperatures in thin polymer films. While it has been shown that the free-surface-induced alteration of a dynamical length scale can generate an enhanced surface mobility as compared to the bulk, as well as a $T_\text{g}$ reduction~\cite{forrest2014does, salez2015cooperative}, there is a persistent debate around the possible importance of the non-equilibrium nature of glassy films made from spin-coated polymers \cite{panagopoulou2017irreversible}. The underlying idea is that such spin-coated PS films exhibit adsorbed layers with strongly out-of-equilibrium extended chain conformations, and that these layers can have a dominant effect on the measured $T_\text{g}$. Besides, a previous work \cite{qi2008substrate} has shown that surface dynamics can be affected by the interactions and dynamics near the substrate, which, when combined to the non-equilibrium scenario above, would then imply that the free surfaces of freshly-cooled and near-equilibrium supported glassy films would show significant differences in their dynamics. However, the current work shows that the dynamical surface properties of freshly-cooled glasses are the same as the ones of vapour-deposited stable glasses -- equivalent to materials aged for $\sim$ 300 years -- and exhibit strong similarities to those of spin-coated high-$M_\text{w}$ PS films.

In summary, we have made a detailed experimental and theoretical study of the bulk and surface dynamics of glassy polystyrene films produced by vapour deposition. In particular, we have compared the surface relaxation times and mobilities of as-deposited (\textit{i.e.} stable) glasses to those of rejuvenated (\textit{i.e.} fresh) glasses. In all cases, shorter surface relaxation times and larger surface mobilities were observed as compared to the bulk values, and the measured surface relaxation times were indistinguishable between the two types of samples. The temperature dependencies of these surface properties were also systematically measured over an extended temperature range, and were found to be mostly indistinguishable between as-deposited and rejuvenated glasses. Finally, strong similarities with the surface behaviours of high-$M_\text{w}$ PS films prepared by spin-coating, were revealed.

\section*{Acknowledgments}
The authors would like to acknowledge many helpful discussions with  Z. Fakhraai. Financial support from Natural Sciences and Research Council of Canada is gratefully acknowledged.
The authors also acknowledge financial support from the European Union through the European Research Council under EMetBrown (ERC-CoG-101039103) grant. Views and opinions expressed are however those of the authors only and do not necessarily reflect those of the European Union or the European Research Council. Neither the European Union nor the granting authority can be held responsible for them. In addition, the authors acknowledge financial support from the Agence Nationale de la Recherche under EMetBrown (ANR-21-ERCC-0010-01), Softer (ANR-21-CE06-0029) and Fricolas (ANR-21-CE06-0039) grants. Finally, they thank the Soft Matter Collaborative Research Unit, Frontier Research Center for Advanced Material and Life Science, Faculty of Advanced Life Science at Hokkaido University, Sapporo, Japan. 
\bibliographystyle{naturemag}
\bibliography{main}
\end{document}


\title{Supplementary Information for \\ Surface and Bulk Relaxation of Vapour-Deposited Polystyrene Glasses}
	\author[1]{Junjie Yin}
	\author[2]{Christian Pedersen}
	\author[1]{Michael F. Thees}
	\author[2]{Andreas Carlson}
	\author[3]{Thomas Salez}
	\author[1]{James A. Forrest}
	\affil[1]{Department of Physics \& Astronomy, University of Waterloo, 200 University Ave. W, Waterloo, ON, N2L 3G1, Canada }
	\affil[2]{Mechanics Division, Department of Mathematics, University of Oslo, 0316 Oslo, Norway}
	\affil[3]{Univ. Bordeaux, CNRS, LOMA, UMR 5798, F-33400, Talence, France}
	\maketitle
	
\begin{figure*}
	\centering
	\includegraphics[width=0.9\textwidth]{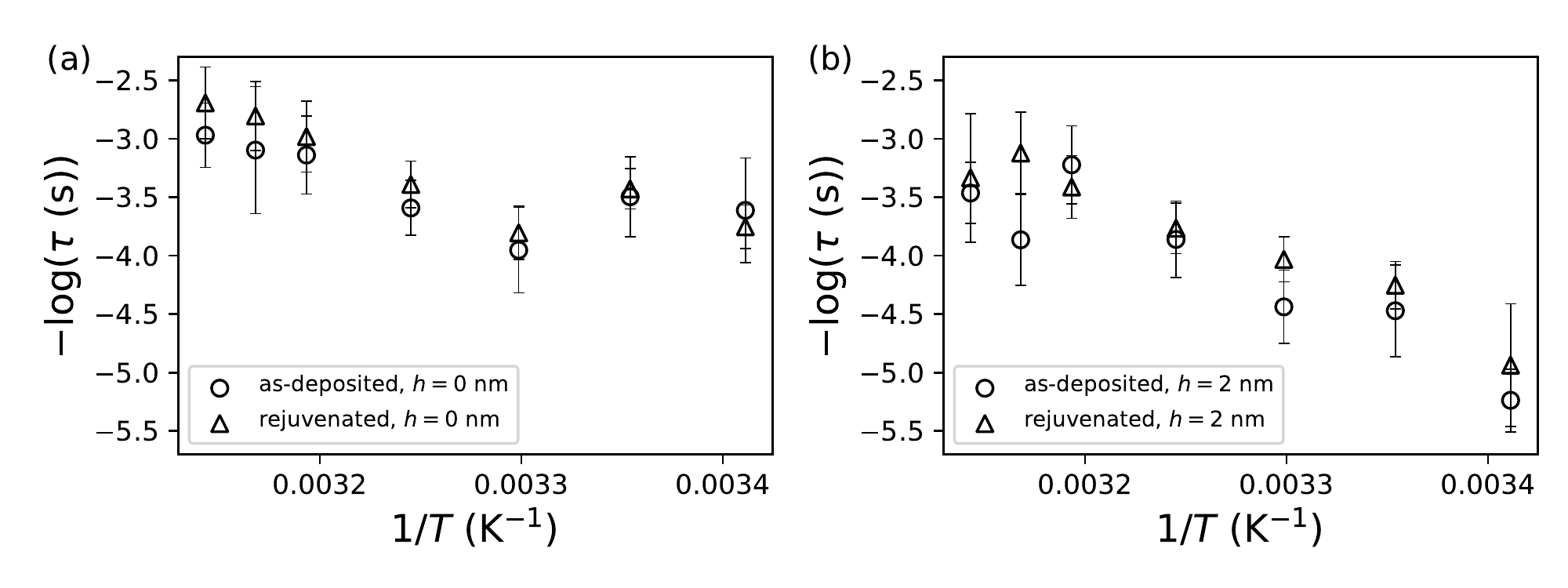}
	\caption{The doubling time of a peak width $d^*$ defined in two different ways. In panel (a) $d^*$ is defined as the minimal value of the radial coordinate $r$ at which the height crosses 0 nm, while in panel (b) it is defined as the minimal value of the radial coordinate $r$ at which the height crosses 2 nm.}
	\label{figS1}
\end{figure*}

From Fig. 8 in the main manuscript, it is observed that the temperature dependence of the experimentally defined surface relaxation time $\tau$ appears to be different than that of the surface mobility obtained from the best fits to the numerical solutions of the GTFEN model. From high temperatures to above  $1/T \sim 3.25 \times 10^{-3}\ \mathrm{K^{-1}}$, the relaxation time $\tau$ transforms from an Arrhenius behaviour to an athermal one, while the same saturation is absent in the surface mobility $M$. Similarly Fig. 8(a) in the main manuscript shows no differences between surface relaxation times of as-deposited and rejuvenated films whereas the mobility shows a small but clear difference. 

These differences appear to be due to the choice of the physical quantity used in characterizing the surface relaxation time. While the doubling time of a specific $d^*$ is a very convenient ways to quantify the time scale of the surface relaxation, the choice of $d^*$ itself can lead to quantitatively and sometimes even qualitatively different results. For example, both Fig.~\ref{figS1}(a) and Fig.~\ref{figS1}(b) show the doubling time of some width $d^*$ defined from the surface profile. In Fig.~\ref{figS1}(a) the definition of $d^*$ is exactly the same as presented in Fig. 8(a) in the main manuscript, which is the minimal value of the radial coordinate $r$ at which the height crosses 0 nm, while in Fig.~\ref{figS1}(b) $d^*$  is chosen as the minimal value of the radial coordinate $r$ at which the height crosses 2 nm (such as that used in ref ~\cite{zhang2017invariant}). It is readily seen that a shift of 2 nm in the observed location changes the temperature dependence of $\tau$ significantly. While a $d^*$ at 0 nm leads to an order of magnitude of change in $\tau$ within the temperature range investigated, a $d^*$ at 2 nm gives almost two orders of magnitude of change in $\tau$. The transition from an Arrhenius to an athermal behaviour is also only present in one case and not the other. Similarly, while still within quoted uncertainties the as-deposited and rejuvenated samples appear to be less similar when a $d^*$ of 2 nm is chosen. 

\begin{figure*}
	\centering
	\includegraphics[width=0.5\textwidth]{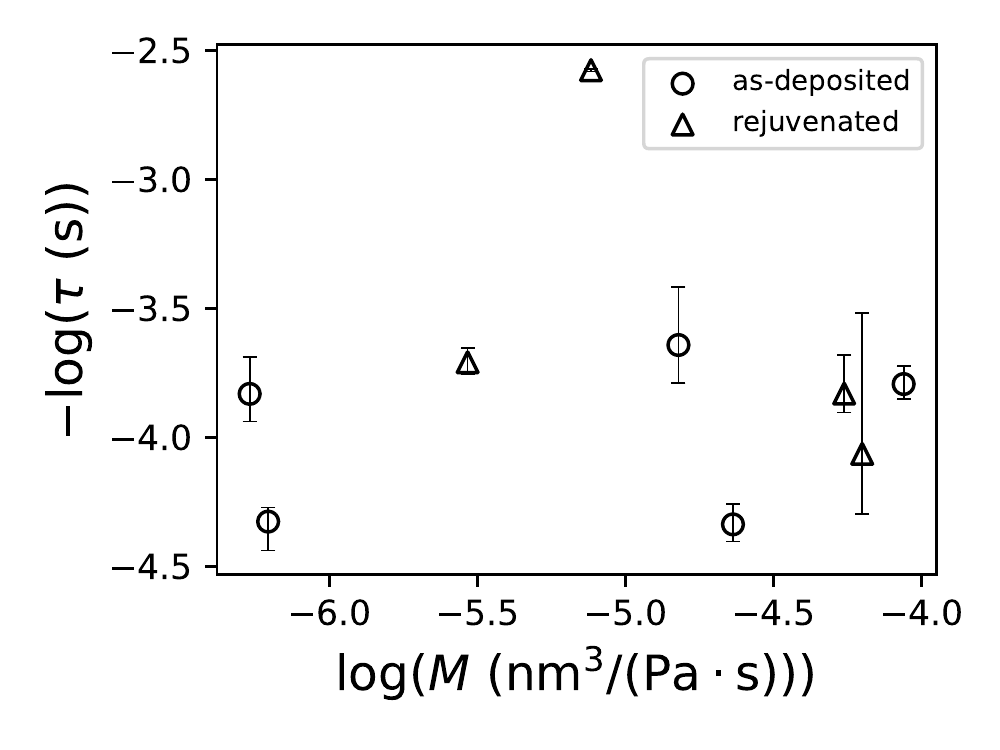}
	\caption{The doubling time of a peak width $d^*$ determined from numerically generated profiles for different values of the mobility $M$}
	\label{figS2}
\end{figure*}

In obtaining the surface mobility $M$, the experimental surface profiles, from $r=0$ to very far from the nanoparticle, are fit to the numerical solutions of the GTFEN model. Unlike the surface mobility $M$ which is a global characteristic of the surface flow, the peak width $d^*$ is a single point, and not every point is equally sensitive to changes in the mobility. For example, in ref \cite{chai2014direct}, it is clear that not every point could be used to determine mobility, and the midpoint of the step in that case is in fact a fixed point and completely insensitive to changes in mobility. In general, the globally determined $M$ value is a more reliable and sensitive way to characterize the surface property, while the report from any locally defined doubling time is less sensitive to changes in mobility and also appear to  depend strongly  on the geometry and the choice of  $d^*$. As a test of this idea we used the profiles generated from the GTFEN model to generate a $d^*$ as a function of time and plotted that doubling time as a function of $M$. This exercise, the results of which are displayed in Fig.~\ref{figS2} revealed that as mobility changes by two orders of magnitude, the times derived from $d^*$ values determined from numerically generated profiles did not similarly change. This demonstrates conclusively that local measurements can result in loss of sensitivity to changes in mobility. 

This discrepancy is also reflected in the literature investigating surface properties of glassy  films using different methods. In the study of the relaxation of nanoholes on PS thin films \cite{fakhraai2008measuring}, the depth of holes are measured and a relaxation time $\tau$ is extracted from their time evolution. A levelling-off in the temperature dependence of $\tau$ at low temperatures is observed, similar to that of $\tau$ in the current study, and even their transition temperatures are strikingly similar. In contrast, in the stepped PS films study by Chai \textit{et al} \cite{chai2014direct} and in the study of surface evolution of PS films upon annealing by Yang \textit{et al} \cite{yang2010glass}, an Arrhenius behaviour in the surface mobility is observed in both studies, similar to that of $M$ in the current study. It is worth noting that in all of the latter three studies, a mathematical model is built, starting from the Stokes equation, to extract the mobility from experimental surface profiles which cover a wide range of the surface. This apparent disagreement between $\tau$ and $M$ can be easily understood with the argument above. When the surface property is described by the time dependence of the profile at a local point, it is influenced by the choice of the point and may not be a good representation of the global relaxation.

\bibliographystyle{naturemag}
\bibliography{supp}